\documentclass{aa}  

\makeatletter
\renewcommand*\aa@pageof{, page \thepage{} of \pageref*{LastPage}}
\makeatother

\usepackage{graphicx}
\usepackage{newtxtext}
\usepackage[smallerops,cmbraces]{newtxmath}
\usepackage{xcolor}
\definecolor{xlinkcolor}{cmyk}{1,1,0,0}
\usepackage[
  bookmarks=true,         
  pdfnewwindow=true,      
  colorlinks=true,        
  linkcolor=xlinkcolor,   
  citecolor=xlinkcolor,   
  filecolor=xlinkcolor,   
  urlcolor=xlinkcolor,    
  final=true,
]{hyperref}
\usepackage{soul}

\usepackage{bm} 
\usepackage{textgreek} 

\usepackage[capitalize]{cleveref} 
\crefname{section}{Sect.}{Sects.}
\creflabelformat{equation}{#2#1#3} 
\crefname{enumi}{item}{items} 

\usepackage{siunitx} 
\DeclareSIUnit[number-unit-product = ]\percent{\char`\%} 

\usepackage[normalem]{ulem} 

\newcommand*{\name}[1]{\textsc{#1}} 

\newcommand*{\gadget}[1]{\name{Gadget-#1}}
\newcommand*{\subfind}{\name{SubFind}}
\newcommand*{\sauron}{SAURON}
\newcommand*{\atlas}{ATLAS\textsuperscript{3D}}
\newcommand*{\matlas}{MATLAS}

\newcommand*{\MagneticumBox}[2]{Box#1 (#2)}

\DeclareSIUnit\parsec{pc}
\DeclareSIUnit\dex{dex}
\DeclareSIUnit\h{\mathnormal{h}}
\DeclareSIUnit\year{yr}
\DeclareSIUnit\years{yrs}
\DeclareSIUnit\arcsec{arcsec}
\DeclareSIUnit\arcmin{arcmin}
\DeclareSIUnit\Msun{M_\odot}
\DeclareSIUnit\Rsun{R_\odot}
\DeclareSIUnit\Lsun{L_\odot}
\DeclareSIUnit\Mvir{\mathnormal{M}_\mathrm{vir}}
\DeclareSIUnit\Rvir{\mathnormal{R}_\mathrm{vir}}
\DeclareSIUnit\Rhalf{\mathnormal{R}_{1/2}}
\DeclareSIUnit\erg{erg}
\DeclareSIUnit\angstrom{\text{Å}}

\newcommand*{\Msun}{\ensuremath{\mathrm{M}_\odot}} 
\newcommand*{\Rsun}{\ensuremath{\mathrm{R}_\odot}} 
\newcommand*{\Lsun}{\ensuremath{\mathrm{L}_\odot}} 
\newcommand*{\Mvir}{\ensuremath{M_\mathrm{vir}}} 
\newcommand*{\Rvir}{\ensuremath{R_\mathrm{vir}}} 
\newcommand*{\Rhalf}{\ensuremath{R_{1/2}}} 

\usepackage{calc}
\newsavebox\CBox
\newcommand\scancel[2][0.5pt]{%
  \ifmmode\sbox\CBox{$#2$}\else\sbox\CBox{#2}\fi%
  \makebox[0pt][l]{\usebox\CBox}%
  \rule[0.5\ht\CBox-#1/2]{\wd\CBox}{#1}}

\begin{document}

\title{A stream come true: Connecting tidal tails, shells, streams,\\ and planes with galaxy kinematics and formation history}
\titlerunning{Tidal features, kinematics, and galaxy formation history}

\author{
    Lucas M.\ Valenzuela\inst{\ref{inst:usm}}
    \and
    Rhea-Silvia Remus\inst{\ref{inst:usm}}
}
\authorrunning{L.\ M.\ Valenzuela \& R.-S.\ Remus}

\institute{
    Universitäts-Sternwarte, Fakultät für Physik, Ludwig-Maximilians-Universität München, Scheinerstr.\ 1, 81679 München, Germany\label{inst:usm}\\
    \email{lval@usm.lmu.de}
}

\date{Received XXX / Accepted YYY}

\abstract
{The rapidly improving quality and resolution of both low surface brightness observations and cosmological simulations of galaxies enable us to address the important question of how the formation history is imprinted in the outer unrelaxed regions of galaxies, and to inspect the correlations of these imprints with another tracer of galaxy formation, the internal kinematics.}
{Using the hydrodynamical cosmological simulation called Magneticum Pathfinder, we identified tidal tails, shells, streams, and satellite planes, and connected them to the amount of rotational support and the formation histories of the host galaxies. This presents the first combined statistical census considering all these four types of features in hydrodynamical cosmological simulations.}
{Tidal features were visually classified from a three-dimensional rendering of the simulated galaxies by several scientists independently. Only features that were identified by at least half of the participating individuals were considered to be existing features. The data on satellite planes and kinematic properties of the simulated galaxies were taken from previous work. The results were compared to observations, especially from the \matlas{} survey.}
{Generally, prominent features are much more common around elliptical than around disk galaxies. Shells are preferentially found around kinematically slowly rotating galaxies in both simulations and observations, while streams can be found around all types of galaxies, with a slightly higher probability to be present around less rotationally supported galaxies. Tails and satellite planes, however, appear independently of the internal kinematics of the central galaxy, indicating that they are formed through processes that have not (yet) affected the internal kinematics. Prolate rotators have the overall highest probability to exhibit tidal features, but the highest likelihood for a specific type of feature is found for galaxies with kinematically distinct cores (KDCs), nearly \SI{20}{\percent} of which exhibit streams.}
{As shells are formed through radial merger events while streams are remnants of circular merger infall, this suggests that the orbital angular momentum of the merger event plays a more crucial role in transforming the host galaxy than previously anticipated. The existence of a shell around a given slow rotator furthermore is a sign of a radial merger formation for this particular slow rotator because one-third of the galaxies with a shell were transformed into slow rotators by the merger event that also caused the shells. The appearance of a stream around a KDC is a direct indicator for the multiple merger formation pathway of that KDC as opposed to the major merger pathway.}


\maketitle
%

\section{Introduction}
\label{sec:introduction}

The outer stellar regions of galaxies encode clues about their formation histories because the relaxation times are longer. Because these outskirts have a low surface brightness (LSB), deep imaging is necessary to detect faint structures that are created through tidal interactions and mergers with other galaxies and that therefore store information about the merger history \citep[e.g.,][]{johnston+08}. A central question in galaxy physics is how details about the formation history of a galaxy may be inferred through its present-day properties. Finding connections between these tidal features and the internal properties of a galaxy, which are less difficult to observe, and connecting them to different aspects of the formation history therefore is an important step toward deciphering galaxy evolution.
These tidal features, that is, shells, streams, or tidal tails, have recently been examined more closely through deep observations \citep[e.g.,][]{bilek+20} and high-resolution simulations \citep[e.g.,][]{karademir+19} with the aim of better understanding their connection with the merger histories of galaxies.

To study tidal features in a more statistically significant way, LSB features have been systematically classified in galaxy surveys such as \matlas{} \citep{duc+15:atlas3dXXIX,bilek+20,sola+22} or the Stellar Stream Legacy Survey \citep{martinez_delgado+23}. The features are often identified through visual inspection, but fully automated approaches have been developed as well. However, these automated methods sometimes have problems that are created through their sensitivity to LSB features \citep{sola+22}.

Methods of identifying tidal features have also been applied to simulations. The possibility of tracing back individual galaxies and their tidal features through time enables us to directly connect the LSB structures to the formation histories of galaxies: Shells typically form through mergers on radial orbits and streams through mergers on circular orbits \citep[e.g.,][]{amorisco15,hendel&johnston15,pop+18,karademir+19}. Investigating the survival timescales in addition to the origin of different tidal features, \citet{mancillas+19} found that shells and streams survive longest and tails are the shortest-lived features. Studies of the appearance of certain tidal features have also been performed in hydrodynamical cosmological simulations, such as for shells by \citet{pop+18}, who found that more massive galaxies are more likely to have shells around them and that shells are typically created through infalling satellites with low orbital angular momenta. \Citet{blumenthal+20} studied the occurrence of tidal features in simulations in the context of interacting galaxies and reported that tidal features and disturbed morphologies are more likely to be observed around interacting pairs of galaxies. \citet{vera_casanova+22} found that the brightest stellar streams formed from early accretion events.

While tidal features typically indicate ongoing or past encounters and mergers with other galaxies, many galaxies still reside in the outskirts of the host galaxy, forming its satellite population.
These are expected to eventually merge with the central galaxy in the future, potentially forming tidal features in the process. Therefore, these satellites are reservoirs for future tidal features.
In this paper, we are also interested in the question whether the appearance of tidal features is influenced by the infall of satellite galaxies being isotropic or anisotropic, that is, whether they are more common around galaxies whose satellites are aligned in planes.
\Citet{lynden_bell76} already found that the Magellanic Clouds and a number of globular clusters form a thin plane around the Milky Way. More satellites have been identified to lie within this plane since then \citep[e.g.,][]{kroupa+05}. Other well-known systems with satellite planes are the Andromeda galaxy \citep[e.g.,][]{ibata+13} and Centaurus~A \citep[e.g.,][]{mueller+18}.
More recently, \citet{heesters+21} found that about \SI{25}{\percent} of \matlas{} galaxies feature a flattened satellite plane in their outskirts.

Satellite planes have been studied in cosmological simulations as well: Some studies reported that very few systems have planes like this \citep[e.g.,][]{pawlowski+14}, whereas others found a better agreement with observations \citep[e.g.,][]{wang+13,foerster+22}.
Simulations furthermore suggest that the arrangement of satellites is connected to the surrounding cosmic web \citep[e.g.,][]{welker+18}, or, like tidal features, may also be the result of galaxy interactions and mergers \citep[e.g.,][]{pawlowski+11,smith+16,bilek+18}. Some studies have concluded that thin planes of satellite galaxies are short-lived \citep{fernando+17}. However, the connection of these satellite planes to the central galaxy and their tidal features is still debated.

A direct imprint of the formation history of a galaxy is also found in its kinematics. While late-type galaxies (LTGs) usually exhibit regular rotation patterns, early-type galaxies (ETGs) show a wide variety of rotational properties. A kinematic classification of ETGs was introduced by \citet{emsellem+07:sauronIX,emsellem+11:atlas3dIII} for the \sauron{} and \atlas{} surveys, where ETGs are split into fast and slow rotators, depending on how rotationally supported they are.
These rotation patterns are closely related to the formation of the galaxy, and slow rotators usually indicate a more violent accretion history \citep[e.g.,][]{schulze+18,schulze+20,lagos+22}. Moreover, special kinematic features such as kinematically distinct cores (KDCs) have been shown to indicate specific formation pathways. For example, large, old KDCs are formed through minor and mini mergers, in which only the inner rotating old disk is preserved \citep{schulze+20}, while small, young KDCs consist of stars that formed during major mergers with gas fractions of 15--\SI{40}{\percent} \citep{hoffman+10} that occurred within the last 3--\SI{4}{\giga\year} \citep{schulze+17}. Prolate rotation patterns have been discussed to originate from overly excessive merging, often through radial orbits, but no clear singular formation pathway has been established for these galaxies so far \citep{ebrova&lokas17,hegde+22}.

As both the internal kinematics and the faint structures found in the outskirts of galaxies are remnants of their formation history, we investigate the correlation between these two tracers in simulations and observations in this work. We consider features related to interactions and mergers with other galaxies and their remnants, that is, tidal features, and also satellite planes. We first introduce the simulations and observations in \cref{sec:sims_obs}, followed by the details of our methods and the definition of the relevant properties in \cref{sec:methods}. Finally, the results are presented and discussed in \cref{sec:results}, and we conclude in \cref{sec:conclusion}.

\section{Simulations and observations}
\label{sec:sims_obs}

\subsection{The Magneticum Pathfinder simulations}
\label{sec:magneticum}

The galaxies classified in this work were extracted from the Magneticum Pathfinder\footnote{\url{www.magneticum.org}} simulation suite (Dolag et al., in prep.), which consists of hydrodynamical cosmological simulations performed with \gadget{3}, an extended version of \gadget{2} (\citealp{springel05:gadget2}; for details of the implementations, see \citealp{teklu+15}).
In particular, we used \MagneticumBox{4}{uhr} (side length: \SI{68}{\mega\parsec}, with a stellar particle mass $\langle m_*\rangle= \SI{1.3e6}{\Msun\per\h}$ and a softening length $\epsilon_* = \SI{1}{\kilo\parsec}$). Galaxies were identified using a modified version of \subfind{} adapted for baryonic matter \citep{springel+01:subfind,dolag+09:subfind}.

The galaxies from Magneticum \MagneticumBox{4}{uhr} have been shown to agree well with observations, including their angular momenta \citep{teklu+15}, kinematics \citep{schulze+18,van_de_sande+19,schulze+20}, dynamics \citep{remus+17,teklu+17,harris+20}, and their in situ component fractions \citep{remus&forbes22}.

In this work, we included all main galaxies with virial masses $\Mvir \geq \SI{7.1e11}{\Msun}$ and stellar masses $M_* \geq \SI{2.4e10}{\Msun}$ to ensure a sufficient resolution for identifying stellar tidal features. Additionally, we limited the sample to galaxies with stellar half-mass radii larger than \SI{2}{\kilo\parsec}, that is, twice the stellar softening length (following \citealp{schulze+18}).
The resulting sample consists of 520~galaxies.
Their ellipticities and $\lambda_{R_e}$-values (a 2D projected proxy for the angular momentum within the effective radius, $R_e$; see \cref{sec:rotators} for more details) at one half-mass radius were taken from \citet{schulze+18} and the satellite plane properties were adopted from \citet{foerster+22}. For more details of these properties, we refer to the original papers.

\subsection{The \matlas{} sample}
\label{sec:matlas}

The observational comparison sample was taken from the \matlas{} survey (introduction by \citealp{duc+15:atlas3dXXIX}; the full sample was presented by \citealp{bilek+20}; the dwarf galaxy sample was presented by \citealp{habas+20:matlas}), which includes 177 massive ETGs with deep optical images. The \matlas{} sample is part of the \atlas{} sample \citep{cappellari+11:atlas3dI}, which comprises 260 nearby galaxies (${<}\SI{42}{\mega\parsec}$) as a complete volume-limited sample. The \matlas{} sample contains all \atlas{} galaxies that do not belong to the Virgo cluster and are not located near bright stars.

Specifically, we used the results obtained by \citet{bilek+20} for the tidal features around the 177 \matlas{} galaxies, where tidal shells, streams, and tails were visually identified, and we used the identifications of satellite planes from \citet[][]{heesters+21}. The ellipticities and $\lambda_{R_e}$ values at one effective radius were taken from \citet{emsellem+11:atlas3dIII}.

\section{Methods and definitions}
\label{sec:methods}

For the comparison between the simulated and observed galaxies, we introduce in the following the different features and properties we used. These include the tidal features we classified, the method with which the satellite planes were identified, and the distinction between fast and slow rotators.

\subsection{Tidal features: Shells, streams, and tidal arms}
\label{sec:classification}

\begin{figure}
    \centering
    \includegraphics[width=0.85\columnwidth]{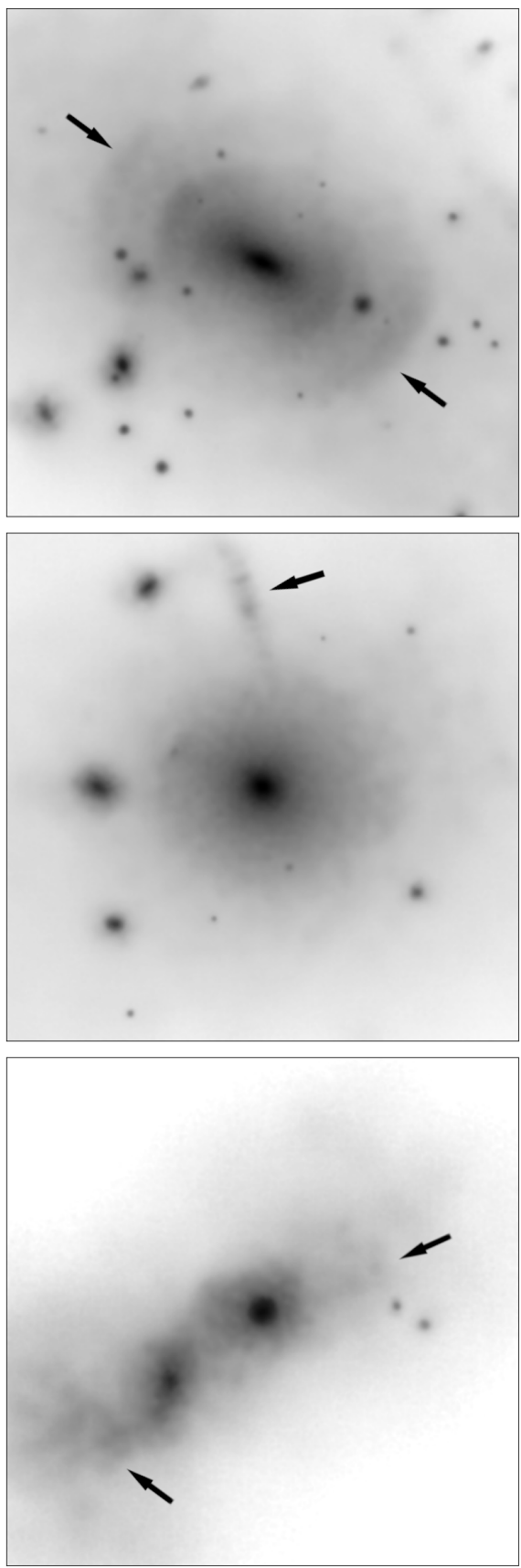}
    \caption{Mock images of simulated galaxies from Magneticum Box4 (uhr) with prominent tidal features: Shells (top; the outer two shells are marked), a stream (middle), and tails (bottom). The mock images were created with the software from \citealp{martin+22} for the r band.}
    \label{fig:mocks}
\end{figure}

The three types of tidal features we considered are shells, streams, and tails. Examples of each feature are shown in \cref{fig:mocks}. They indicate past or ongoing gravitational interactions with other galaxies. Shells appear as circular arcs centered around a galaxy (top panel of \cref{fig:mocks}). They have sharp outer borders and are thought to form through the radial infall and disruption of (massive) satellites \citep[e.g.,][]{quinn84,pop+18,karademir+19,bilek+22}. Streams are thin slivers of stars around the host galaxy (middle panel of \cref{fig:mocks}) that are formed through the tidal disruption of an infalling galaxy on a circular orbit \citep[e.g.,][]{karademir+19,tutukov+21}. Tails are stellar arms that are stripped from the host galaxy (bottom panel of \cref{fig:mocks}). They are the result of tidal interaction with another galaxy, where the interaction is either ongoing or occurred recently, up to \SI{2}{\giga\year} ago \citep{mancillas+19}.

We classified the galaxies in Magneticum with respect to the existence of these tidal features by visual inspection of the 3D stellar component to find any sufficiently prominent features independent of their orientation and possible projection effects. Similar to the classification of \citet{bilek+20} for the \matlas{} sample, many people visually classified the galaxies, and no detection algorithm was used. A key difference between the classification methods is that the simulated galaxies were viewed in 3D, whereas the observations of the \matlas{} galaxies only provide data for one projection. We opted for this approach to use the full potential of the simulation and find the actually existing tidal features around the simulated galaxies to obtain results independent of projection because \citet{pop+18} found, for example, that the shells of only \SI{40}{\percent} of the galaxies with shells are visible in all three considered projections, whereas most shells are only visible in two projections. By performing the classification in 3D, we aimed to find all tidal features, which enabled us to investigate the underlying connection of the features with the formation history and especially galaxy kinematics. The comparison to observations is intended to determine whether the same trends are found in observations. The comparison is  not intended as a verification of the simulation. A verification like this has already been shown in previous studies, especially for the kinematics \citep{teklu+15,schulze+18,schulze+20}.

Seven scientists (two of which are experienced scientists in galaxies and their dynamics, while the other five were instructed through observed examples about the look of the different tidal features) participated in the classification, for which they were presented with a 3D rendering of a given galaxy that was rotated to be viewed from all sides. The local environment around the individual galaxies was visible as well. The classification was performed independently by the participants. Each person stated whether they identified any tidal shells, streams, and tails, or no features. Analysis of the classifications did not result in a significant difference between the classifications of the experienced and less experienced scientists. One difference to point out is that the more experienced scientists found that more galaxies had streams. However, this difference was negligibly small.

In the following, we consider a galaxy to have a given tidal feature if at least half of the participants were in favor of this. For the \matlas{} classification by \citet{bilek+20}, this is considered to be the case if the classification score is at least~1 (the maximum score for them is~2, where 1 means that the respective feature is likely to be present).

We did not emulate a surface brightness limit cut for the simulations in this study. We refrained from this approach for the following reason: First, we would have had to introduce a model to obtain luminosities from the simulated masses, which has its own set of uncertainties, especially as we wished to make the comparison based on the true 3D findings from the simulations, as discussed above.
Furthermore, the intrinsic mass resolution of the simulation results in a surface brightness limit that by chance is similar to that of the \matlas{} survey when a constant mass-to-light ratio is assumed (see also \citealp{remus&forbes22} for more details of surface brightness comparisons for the simulation).
While this should be kept in mind, we therefore do not consider the comparability to be an issue given the fortunate similarity between the simulation resolution and the surface brightness limits of the \matlas{} survey.

\subsection{Satellite planes}
\label{sec:planes}

We considered a galaxy to have a satellite plane when at least \SI{80}{\percent} of its satellite galaxies build a plane with a thickness ${\leq}\SI{0.2}{\Rvir}$ and when the in-plane angular momentum makes up at least \SI{90}{\percent} of their total angular momentum. These properties were computed through the momentum in the thinnest plane (MTP) method \citep{foerster+22}.

\subsection{Fast and slow rotators, and kinematic classes}
\label{sec:rotators}

To distinguish fast from slowly rotating ETGs, we followed the approach taken by \citet{emsellem+11:atlas3dIII}. They determined an empirical cut in the $\lambda_{R_e}$-$\epsilon_e$ plane, where $\epsilon_e$ is the ellipticity at the effective radius. The $\lambda_{R}$ value was introduced by \citet{emsellem+07:sauronIX} as part of the \sauron{} project, defined as
\begin{equation}
    \lambda_R = \frac{\langle R |V| \rangle}{\big\langle R \sqrt{V^2 + \sigma^2} \big\rangle},
\end{equation}
with the projected radius $R$, the line-of-sight velocity $V$, the line-of-sight velocity dispersion $\sigma$, and the flux-weighted averages $\langle \cdot \rangle$. The adaptation for simulations, where no fluxes but only the masses $M$ are available, requires $\langle \cdot \rangle$ to be the mass-weighted averages over the quantities \citep[e.g.,][]{jesseit+09,naab+14:atlas3dXXV,schulze+18}, assuming a constant mass-to-light ratio (see \citealp{schulze+18} for more details).
As defined by \citet{emsellem+11:atlas3dIII}, galaxies that fulfill $\lambda_{R_e} > 0.31 \sqrt{\epsilon_e}$ are classified as fast rotators, and all other galaxies are classified as slow rotators.

We obtained these parameters for the simulated galaxies from the simulation for the edge-on projections of the galaxies because this maximizes the $\lambda_{R_e}$ values, and thus, the full kinematic properties are best identified. The distribution of $\lambda_R$-$\epsilon$ using random projected values are, however, comparable to observations, as shown by \citet{schulze+18} and \citet{van_de_sande+19}, who not only compared the simulation to observations from SAMI and \atlas{}, but also to other cosmological simulations.

Galaxies were further classified based on their kinematic patterns within the half-mass radius. The classes used for this study are regular rotators, nonrotators, KDCs, and prolate rotators, based on the classifications from \citet{schulze+18}. These classes are similar to those presented by \citet{krajnovic+11:atlas3dII} for \atlas{} galaxies, with the addition of the prolate rotators as an individual class, while the KDCs encompass all galaxies with separate cores independently of the outer rotation pattern.

\subsection{Morphological classes}

The galaxies from Magneticum were subdivided into morphological types using the classification from \citet{teklu+17} based on the $b$ value \citep{teklu+15}, which is a good proxy for distinguishing LTGs from lenticular galaxies and ETGs \citep{romanowsky&fall12,teklu+15}. The parameter is based on the location of the galaxies in the stellar mass-stellar angular momentum plane and has been shown to successfully determine the morphological galaxy types in simulations \citep[e.g.,][]{teklu+17,schulze+18,emami+21}. We distinguished between clear disk galaxies (LTGs), clear elliptical galaxies (ETGs), and intermediate galaxies. The latter mostly contain S0-like galaxies and are thus often associated with ETGs. In this work, we treated them separately because the formation pathways of S0 galaxies can differ from those of typical elliptical galaxies and are still debated.

\section{Results}
\label{sec:results}

\subsection{Feature statistics and mass, morphology, and size}
\label{sec:mass_size}

\begin{figure*}
    \centering
    \includegraphics[width=\textwidth]{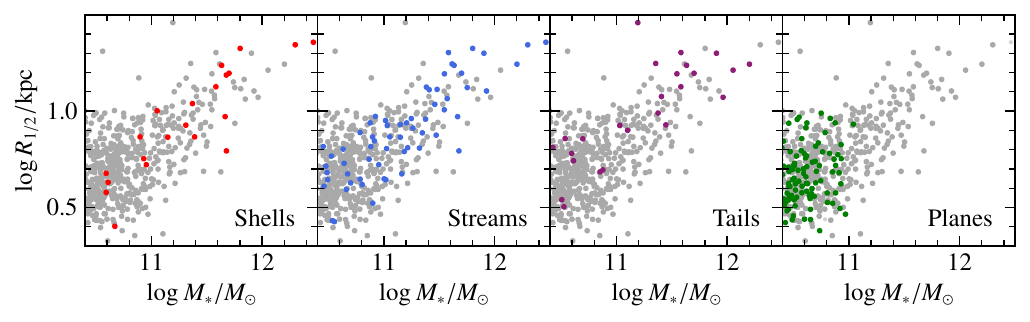}
    \caption{Mass-size relation of the Magneticum galaxy sample. Galaxies are classified as having tidal features or satellite planes in color.}
    \label{fig:mass_size_features}
\end{figure*}

\begin{figure*}
    \centering
    \includegraphics[width=\textwidth]{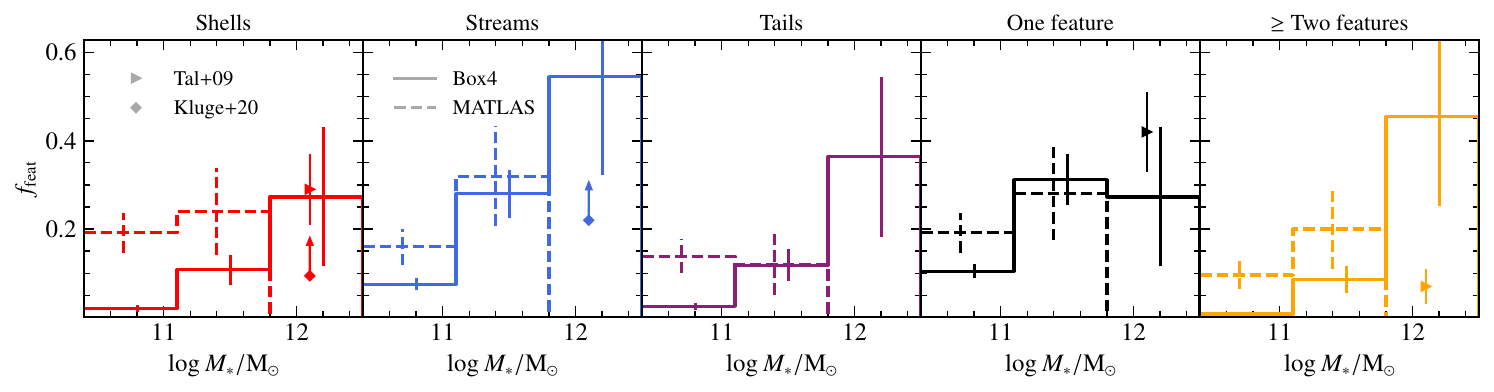}
    \caption{Fraction of galaxies with a given tidal feature (first three panels), any of the three tidal feature types (fourth panel), or a combination of multiple different types of features (fifth panel), $f_\mathrm{feat}$, in three stellar mass bins for Magneticum, \matlas{} \citep{bilek+20}, and for brightest cluster galaxies (BCGs) from \citet{tal+09} and \citet{kluge+20}. The masses for \matlas{} galaxies are taken from the dynamical masses obtained by \citet{cappellari+13:atlas3dXV}, which roughly correspond to the stellar masses. The vertical lines indicate the $1\sigma$ uncertainties.}
    \label{fig:feature_fraction}
\end{figure*}

When the galaxies with a given tidal feature are highlighted in the mass-size relation, all three types of tidal features occur at all sizes and masses, with a clear preference for larger and more massive systems (first three panels of \cref{fig:mass_size_features}), that is, the fraction of galaxies with tidal features increases with stellar mass (\cref{fig:feature_fraction}). The figure also shows that the given mass bins have a larger fraction of galaxies with shells in \matlas{} than in Magneticum, even though the \matlas{} sample does not include the most massive galaxies with $M_* > \SI{5.5e11}{\Msun}$, and thus, the highest-mass bin cannot be compared to \matlas{}. The lowest-mass bin also includes relatively more streams and tails in \matlas{} galaxies. Similarly, relatively more \matlas{} galaxies have any one type or multiple types of tidal feature around them than in Magneticum. To cover the high-mass end, we included the tidal feature measurements around brightest cluster galaxies (BCGs) by \citet{tal+09} and \citet{kluge+20}. The observations of \citet{tal+09} for shells around BCGs are consistent with the simulated fraction in the highest-mass bin, while the observations from \citet{kluge+20} are lower bounds for the fractions of shell and stream galaxies because of the large distance to their observed BCGs \citep[see][]{kluge+20}. Simulations and observations both indicate that tidal features are generally more common around more massive galaxies, and multiple features are the most common around BCGs.

When only galaxies with stellar masses $M_* \geq \SI{e11}{\Msun}$ are considered, about \SI{10}{\percent} to \SI{30}{\percent} have a given tidal feature (see \cref{tab:statistics}). Streams are by far the most common of these tidal features, while shells and tails are found with similar frequencies. The occurrence of shells in \SI{11\pm3}{\percent} of the Magneticum galaxies agrees within the uncertainty limits with \SI{12}{\percent} from \citet{atkinson+13}, who used a different stellar mass range and somewhat different categories (see the discussion by \citealp{bilek+20}), and \SI{17\pm7}{\percent} from \citet{bilek+20} for \matlas{} (for the same stellar mass cut of \SI{e11}{\Msun}), but the uncertainty range does not reach the lower limit of \SI{18\pm3}{\percent} from \citet{pop+18} for the Illustris simulation. The fraction of galaxies with streams in Magneticum (\SI{29\pm5}{\percent}) is consistent with the \SI{25\pm9}{\percent} found for \matlas{}. Finally, the fraction of galaxies with tails in Magneticum (\SI{12\pm4}{\percent}) is larger than that in \matlas{} (\SI{6\pm4}{\percent}), although the values are consistent within the uncertainty limits.
Overall, the abundance of stellar tidal features found in the Magneticum Box4 simulation for the galaxies with higher stellar masses is consistent with previous findings.

\begin{table}
    \centering
    \caption{Number and fraction of galaxies with a type of tidal feature.}
    \label{tab:statistics}
    \begin{tabular}{lcc}
        \hline\hline
        Feature & Number & Fraction \\
        \hline
        Shells  & $14$ & \SI{11\pm3}{\percent} \\
        Streams & $38$ & \SI{29\pm5}{\percent} \\
        Tails   & $16$ & \SI{12\pm4}{\percent} \\
        \hline
    \end{tabular}
    \tablefoot{Only the galaxies with stellar masses $M_* \geq \SI{e11}{\Msun}$ were considered for the numbers and fractions. The total number of galaxies analyzed in this mass range is~131.}
\end{table}

The numbers of galaxies in each morphological class containing tidal features are listed in \cref{tab:statistics_morphology}. While shells preferentially occur around ETGs, streams are found to be equally likely in all three types, and tails are more likely in intermediate galaxies and LTGs, although the uncertainty ranges overlap in the latter case. Because a number of galaxies feature multiple tidal structures in their outskirts, we also determined the fraction of galaxies that contains at least one type of tidal feature (shells, streams, or tidal arms): \SI{18\pm3}{\percent} of ETGs, \SI{17\pm4}{\percent} of intermediate type galaxies, and \SI{17\pm6}{\percent} of LTGs. Overall, the fractions of LTGs (\SI{2\pm2}{\percent})  are similar, with multiple types of tidal feature as for intermediate-type galaxies (\SI{3\pm2}{\percent}) and ETGs (\SI{3\pm2}{\percent}).

\begin{table}
    \centering
    \caption{Numbers and fractions of ETGs, intermediates (Intm.), or LTGs with shells, streams, tails, or satellite planes.}
    \label{tab:statistics_morphology}
    \begin{tabular}{lccccc}
        \hline\hline
        Type & All & Shells & Streams & Tails & Planes \\
        \hline
        ETGs & 295 & $17$ & $37$ & $9$ & $49$ \\
             &     & \SI{6\pm2}{\percent} & \SI{13\pm3}{\percent} & \SI{3\pm2}{\percent} & \SI{17\pm3}{\percent} \\
        Intm. & 172 & $4$ & $20$ & $12$ & $37$ \\
                &     & \SI{2\pm2}{\percent} & \SI{12\pm3}{\percent} & \SI{7\pm3}{\percent} & \SI{22\pm4}{\percent} \\
        LTGs & 53 & $0$ & $5$ & $5$ & $14$ \\
             &    & \SI{0\pm2}{\percent} & \SI{9\pm5}{\percent} & \SI{9\pm5}{\percent} & \SI{26\pm8}{\percent} \\
        \hline
    \end{tabular}
    \tablefoot{The full classified sample of 520~galaxies is considered for the numbers and fractions. The morphological classes of the galaxies were taken from \citet{teklu+17}.}
\end{table}

For the satellite planes, we find that only galaxies with stellar masses ${\lesssim}\SI{e11}{\Msun}$ are surrounded by thin planes of satellites. This is consistent with the results from \citet{foerster+22}, who found that the more massive systems lack thin planes. We find indications that planes are slightly more common around LTGs (\SI{26\pm8}{\percent}) than ETGs (\SI{17\pm3}{\percent}), with the intermediate galaxies in between (\SI{22\pm4}{\percent}). This does not reflect the findings for tidal features and clearly shows that the appearance of satellite planes is not directly connected to the appearance of tidal features. This shows that tidal features do not depend on the large-scale inflow field, but rather on the orbits and mass ratios of the infalling satellites.

\begin{figure*}
    \centering
    \includegraphics[width=\textwidth]{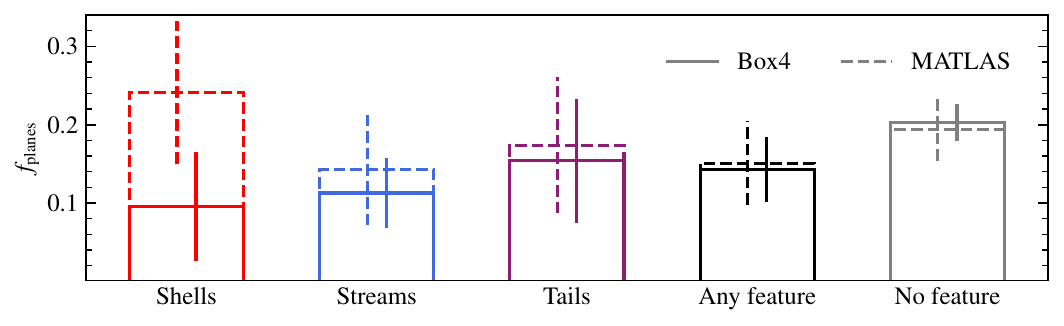}
    \caption{Fraction of galaxies with a satellite plane for different types of galaxies in \MagneticumBox{4}{uhr} \citep{foerster+22} and MATLAS \citep{bilek+20,heesters+21}: galaxies with shells, streams, tails, at least one of these features, and with none of these features. The vertical lines indicate the $1\sigma$ uncertainties.}
    \label{fig:planes_fraction}
\end{figure*}

The fraction of galaxies with a plane of satellites of the galaxies with certain tidal features shows a consistent behavior between the simulation and observations (\cref{fig:planes_fraction}). The largest difference is found for galaxies with shells, where there is an indication that more observed shell galaxies are surrounded by satellite planes. There is also an indication that the fraction of galaxies without features (last vertical bar) surrounded by a satellite plane is larger than for galaxies with tidal features (first four bars). This means that there may be a slight preference for satellite planes to occur around galaxies without tidal features. The same slight trend is also found when we compare the fraction of galaxies with any type of tidal feature for galaxies with satellite planes versus galaxies without satellite planes: \SI{13\pm4}{\percent} of the simulated galaxies (\SI{25\pm9}{\percent} for observed galaxies) with a satellite plane have at least one type of tidal feature, while \SI{19\pm3}{\percent} (\SI{31\pm5}{\percent}) of the galaxies without a satellite plane have at least one type of tidal feature.

\subsection{Correlations of tidal features and satellite planes with kinematic classes}
\label{sec:kinematic_classes}

Using the kinematic classification for the simulated galaxies from \citet{schulze+18} as described in \cref{sec:rotators}, we determined the numbers and fractions of galaxies with the different kinematic classes that have the respective tidal features or a satellite plane (\cref{tab:statistics_kinematic_classes}). In the table, we combined the distinct core (DC) and KDC classes into a single class of KDCs (as also done by \citealp{valenzuela+24}; Remus et al., in prep.).
The trend emerges that regular rotators are less likely to have tidal features, with the exception of tidal tails, which are found at a similar fraction around nonrotators and regular rotators. In contrast, nonrotators and especially prolate rotators have higher overall fractions of shells and streams than regular rotators, although the uncertainty limits overlap in part due to low number statistics.

We again determined the fraction of galaxies that contains at least one type of tidal feature (shells, streams, or tidal arms): \SI{12\pm3}{\percent} of regular rotators, \SI{19\pm5}{\percent} of nonrotators, \SI{27\pm11}{\percent} of KDCs, and \SI{44\pm17}{\percent} of prolate rotators. The much higher fraction of prolate rotators with at least one type of tidal feature compared to any single given feature shows that they are more likely to contain multiple tidal features in their outskirts, whereas the overlap of different types of features around a galaxy is less pronounced for the other kinematic classes.

These results agree with those from \citet{ebrova+21}, who found that a large fraction of observed prolate rotators are surrounded by multiple shells, and all of them show signs of interactions. Their determined fraction of 10 out of 19 systems (\SI{53\pm17}{\percent}) is significantly higher than the 3 out of 16 Magneticum galaxies (\SI{19\pm11}{\percent}). However, when we consider only the simulated prolate galaxies with stellar masses of ${\geq}\SI{e11}{\Msun}$, which all but one of the galaxies considered by \citet{ebrova+21} satisfy (but only 9 of our 16 galaxies), we find that 3 out of 9 galaxies have shells (\SI{33\pm20}{\percent}), which is consistent within the uncertainty limits with the observed fraction of 9 out of 18 from that study (\SI{50\pm17}{\percent}). This reflects the mass trend for shells to be more common around more massive galaxies.

Of all kinematic classes, KDCs show the strongest probability of exhibiting a certain tidal feature, driven by the stream feature. The fact that almost every fourth KDC has streams in its outskirts is consistent with many KDCs having decreasing $\lambda_R$ profiles at radii larger than the half-mass radius, which is usually caused by merger histories dominated by small mergers \citep{schulze+20}. Kinematically distinct cores with these radial $\lambda_R$ profiles are remnants of old disks that have assembled their outskirts by multiple minor mergers on mostly circular orbits \citep{schulze+20}. When we traced the six simulated KDC galaxies with streams back until redshift $z = 1$, all of these galaxies had only grown in mass through mini and minor mergers. Therefore, this connection between the appearance of a stream and a KDC feature is a good indicator that the particular KDC was likely formed through the multiple merger scenario and not through the gas-rich recent major merger scenario, which is the other known formation pathway for KDCs \citep{hoffman+10,schulze+17}.

\begin{table}
    \centering
    \caption{Numbers and fractions of galaxies, with the given kinematic class having shells, streams, tails, or satellite planes.}
    \label{tab:statistics_kinematic_classes}
    \begin{tabular}{lccccc}
        \hline\hline
        Type & All & Shells & Streams & Tails & Planes \\
        \hline
        RR & 286 & $2$ & $26$ & $8$ & $63$ \\
           &     & \SI{1\pm1}{\percent} & \SI{9\pm2}{\percent} & \SI{3\pm1}{\percent} & \SI{22\pm3}{\percent} \\
        NR & 108 & $10$ & $16$ & $2$ & $19$ \\
           &     & \SI{9\pm3}{\percent} & \SI{15\pm4}{\percent} & \SI{2\pm2}{\percent} & \SI{18\pm5}{\percent} \\
        KDC & 26 & $1$ & $6$ & $2$ & $1$ \\
            &    & \SI{4\pm4}{\percent} & \SI{23\pm10}{\percent} & \SI{8\pm6}{\percent} & \SI{4\pm4}{\percent} \\
        PR & 16 & $3$ & $3$ & $2$ & $2$ \\
           &    & \SI{19\pm11}{\percent} & \SI{19\pm11}{\percent} & \SI{13\pm9}{\percent} & \SI{13\pm9}{\percent} \\
        \hline
    \end{tabular}
    \tablefoot{Only the subset of our full sample that was kinematically classified by \citet{schulze+18} was considered for the numbers and fractions. This subsample comprises 436~galaxies. The kinematic classes include regular rotators (RR), nonrotators (NR), KDCs, and prolate rotators (PR).}
\end{table}

Finally, satellite planes again show the opposite behavior of tidal features. They are most common around regular rotators and least common around KDCs and prolate rotators. This shows that the alignment of satellite galaxies in a plane is not connected to the appearance of these features or might possibly even be anticorrelated (as shown in \cref{fig:planes_fraction}).

\subsection{Correlations of tidal features and satellite planes with the inner rotational support}
\label{sec:kinematics}

\begin{figure*}
    \centering
    \includegraphics[width=\textwidth]{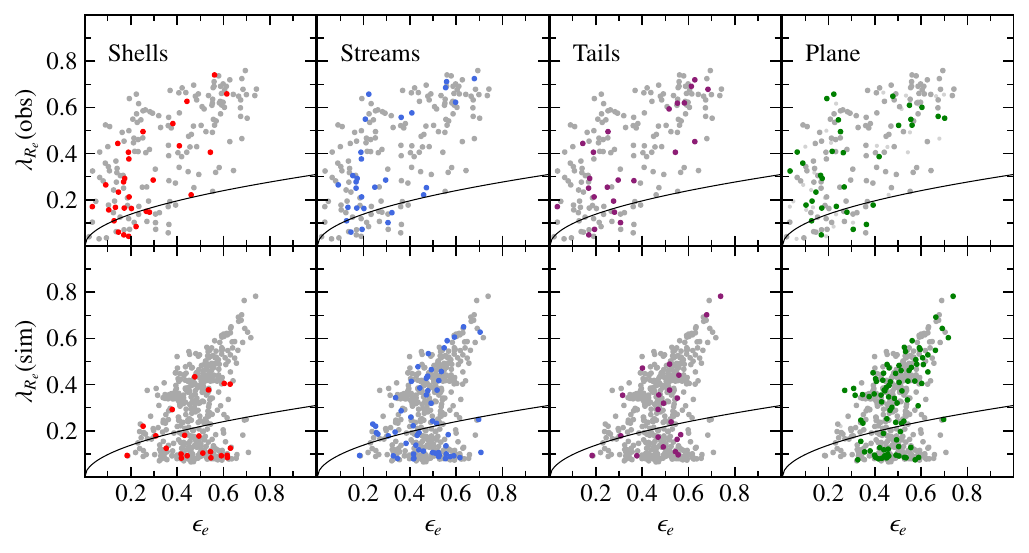}
    \caption{Galaxies classified as having tidal features or a satellite plane in the $\lambda_{R_e}$-$\epsilon_e$ plane in the \matlas{} survey (\emph{top row}) and in Magneticum Box4 (\emph{bottom row}). Only the ETGs and intermediate-type galaxies are shown for the simulated sample. The colored data points represent galaxies with the respective feature. The large gray dots were classified as lacking the respective feature, and small gray dots were not classified (only the case for the observed satellite planes). The values of $\lambda_{R_e}$ and $\epsilon_e$ were determined at one effective radius for the observations and at one half-mass radius for the simulations. For the simulations, they were determined in the edge-on projection. The solid black lines indicate the border between fast and slow rotators \citep{emsellem+11:atlas3dIII}, with fast rotators lying above the border. The measurements of shells, streams, and tails of \matlas{} galaxies are taken from \citet{bilek+20} and the measurements of the planes are taken from \citet{heesters+21}.}
    \label{fig:lambda_r_epsilon_features}
\end{figure*}

We investigated the correlation between two known tracers of the accretion history of a galaxy, the appearance of tidal features, and the internal galaxy kinematics. To measure the latter, we used the position of galaxies in the $\lambda_{R_e}$-$\epsilon_e$ plane for both the observed and simulated galaxies (top and bottom of \cref{fig:lambda_r_epsilon_features}, respectively). The values of $\lambda_{R_e}$ and $\epsilon_e$ of the observed galaxies were obtained from a random projection, while the simulated values are the highest possible values because they are obtained from the edge-on projections (for the projection effects found in this $\lambda_{R_e}$-$\epsilon_e$ space, see also the results from \citealp{cappellari+07:sauronX}). To do this, we limited the simulated sample of galaxies to ETGs and intermediate-type galaxies because the \matlas{} sample does not contain LTGs.
At first glance, there are no clear differences between the distributions of galaxies with tidal features and the total sample for the observations and simulations. The trend of shells lying around slow rotators is immediately visible only for the simulated galaxies.
Similarly, the galaxies with satellite planes are uniformly distributed and do not show any preference for slow or fast rotation, or for particularly round or elongated shapes (right panel of \cref{fig:lambda_r_epsilon_features}).

\begin{figure*}
    \centering
    \includegraphics[width=\textwidth]{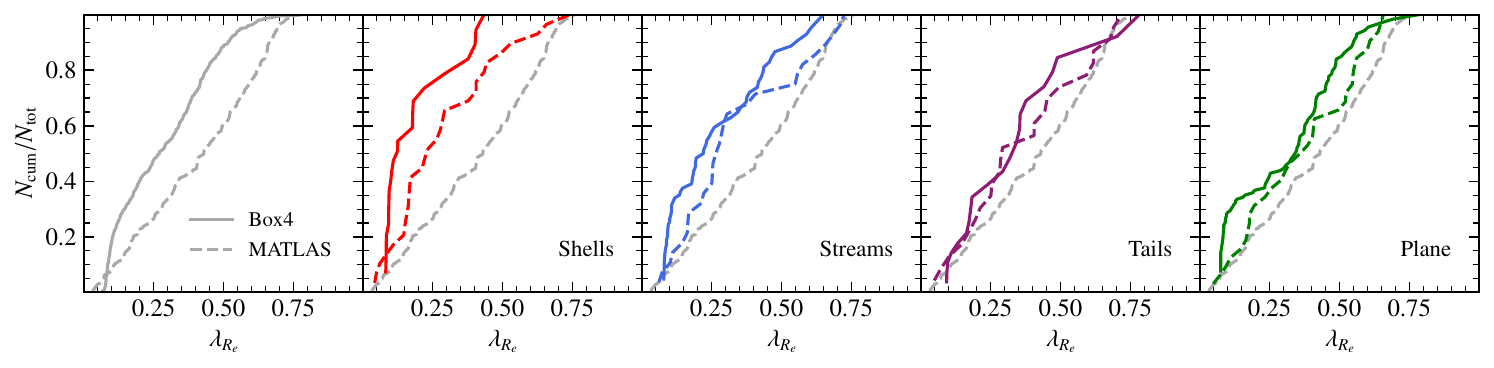}
    \caption{Cumulative histograms of $\lambda_{R_e}$ for the galaxies in the \matlas{} survey (dashed lines) and in Magneticum Box4 (uhr) (solid lines). Only the ETGs and intermediate-type galaxies were considered for the simulated sample. The first plot shows the difference in the distributions between observations and simulations. The cumulative histograms of the simulated galaxies in the other four panels were adapted to match the parent distribution of \matlas{} to be directly comparable to the distributions of the \matlas{} survey. The colored lines lying above the dashed gray line (the overall distribution of the \matlas{} survey) indicate an increased presence of the respective tidal feature or satellite plane for galaxies with lower rotational support compared to those with higher rotational support. The measurements of shells, streams, and tails of \matlas{} galaxies are taken from \citet{bilek+20}, and the measurements of the planes are taken from \citet{heesters+21}.}
    \label{fig:lambda_r_histogram_cum}
\end{figure*}

The differences between the edge-on and random projection of galaxies in the $\lambda_{R_e}$-$\epsilon_e$ plane were discussed in detail by \citet{schulze+18}, especially with respect to the ellipticity, $\epsilon_e$, clearly showing the problems when comparing ellipticities between simulations and observations. In contrast, $\lambda_{R_e}$ is more directly comparable because its values change only slightly for most projections, except when seen nearly face on.
Thus, we determined the cumulative histograms of $\lambda_{R_e}$ for the total simulated and observed galaxy samples, and for the galaxies with the respective tidal features or satellite planes (\cref{fig:lambda_r_histogram_cum}).
Because the $\lambda_{R_e}$ distributions of the full samples are not identical (leftmost panel), a direct comparison between the $\lambda_{R_e}$ distributions of galaxies with a certain tidal feature requires a correction to the values in the cumulative histograms. The method for this correction is described in \cref{app:rescaling}.

By adapting the simulated cumulative histograms to the observed parent distribution, the colored lines in \cref{fig:lambda_r_histogram_cum} are directly comparable with each other. For the following comparisons, we used the two-sample Kolmogorov--Smirnov test with the null hypothesis that the underlying distributions of two random samples are the same. Very similar distributions of $\lambda_{R_e}$ are found for galaxies with a given tidal feature in the simulations and observations ($p$-values\footnote{For the Kolmogorov--Smirnov test, the $p$-value is the probability that two samples were drawn from the same underlying distribution, the null hypothesis. This means that the underlying distributions can be concluded to be different for $p$-values below 0.05 (by convention).} of 0.07, 0.32, 0.91, and 0.23 for shells, streams, tails, and planes, respectively).

For individual tidal features, we find a very clear trend that shells are mostly found around slow rotators, both in the simulation and in observations ($p$-values of 0.014 and 0.008 for Box4 and \matlas{}, respectively, when comparing the tidal feature $\lambda_{R_e}$ distributions to the simulated or observed total sample distributions). Streams have a slight preference to appear around slow and intermediate rotators, but this is only statistically significant here for the observations ($p$-values of 0.23 and 0.02 for Box4 and \matlas{}, respectively). Tidal tails are found around both slow and fast rotators with similar probabilities ($p$-values of 0.84 and 0.39 for Box4 and \matlas{}, respectively). The overall trends agree with the findings of \citet{bilek+23}, who determined a negative correlation between tidal features and rotational support. A direct comparison is not possible, however, because the definitions of rotational support differ.

The same is also true for the satellite planes: Planes are found equally around galaxies with any amount of rotational support in both Magneticum and \matlas{} ($p$-values of 0.09 and 0.59 for Box4 and \matlas{}, respectively). While \citet{heesters+21} stated that \SI{40}{\percent} of their slow rotators featured satellite planes and only \SI{16}{\percent} of the fast rotators do this, they also remarked that these results are not that statistically significant given the small number of slow rotators in the sample. Even though the satellite planes are detected through entirely different methods, with the observed planes being determined from 2D data \citep{heesters+21} and the simulated planes from 3D data, it can be expected that the general correlation of the existence of such planes with the internal kinematics of galaxies would be found to be similar between observations and simulations. This is precisely what we have found between Box4 and MATLAS galaxies.

The result that shells are more common around slow rotators suggests that the processes leading to low rotational support and the appearance of shells are related. Shells have been found to originate from radial merger events, which are more pronounced the more similar the two merging galaxies are in mass \citep[e.g.,][]{amorisco15,pop+18,karademir+19}. Slow rotators are found to often have formed through merger events, in which a large amount of mass has been accreted over a short period of time, most commonly through (major) merger events with mass ratios higher than 1:10 \citep[e.g.,][]{schulze+18,lagos+22}. Therefore, the existence of a shell strongly indicates that the observed slow rotator formed through a recent radial merger event with a mass ratio higher than 1:10. When we trace the formation history of the slow rotators with shells back in the simulation, only 2 out of the 21 simulated galaxies with shells lack mergers above a mass ratio of 1:10 occurring within the past \SI{4}{\giga\year} (the lifetime of shells as estimated by \citealp{mancillas+19}). In seven cases (\SI{33\pm13}{\percent}), the major merger that created the shells also led to a sudden strong decrease in $\lambda_{R_e}$, such that a previous fast rotator was converted into a slow rotator. These slow rotators seem to have particularly late formation times, while the other slow rotators with shells have had a low value of $\lambda_{R_e}$ for more than \SI{4}{\giga\year}, six of which became slow rotators already by $z=2$. This shows that most galaxies in the simulation with shells became slow rotators at early times, in agreement with \citet{bilek+23}, whereas roughly one-third of them just became slow rotators at the same time as the shells were formed. Thus, in all cases where the progenitor galaxy was a fast-rotating galaxy \SI{4}{\giga\year} ago, the major merger that formed the shells also destroyed the inner rotation pattern of the galaxy.

\begin{figure*}
    \centering
    \includegraphics[width=\textwidth]{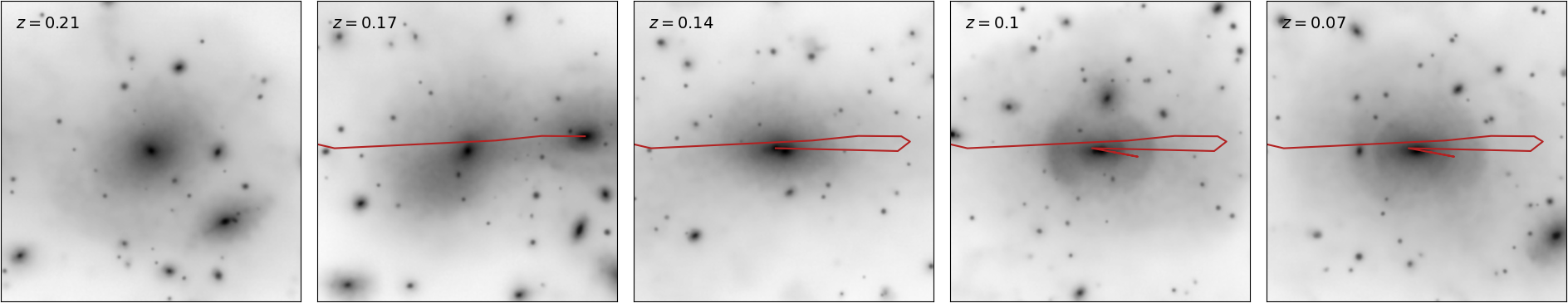}
    \includegraphics[width=\textwidth]{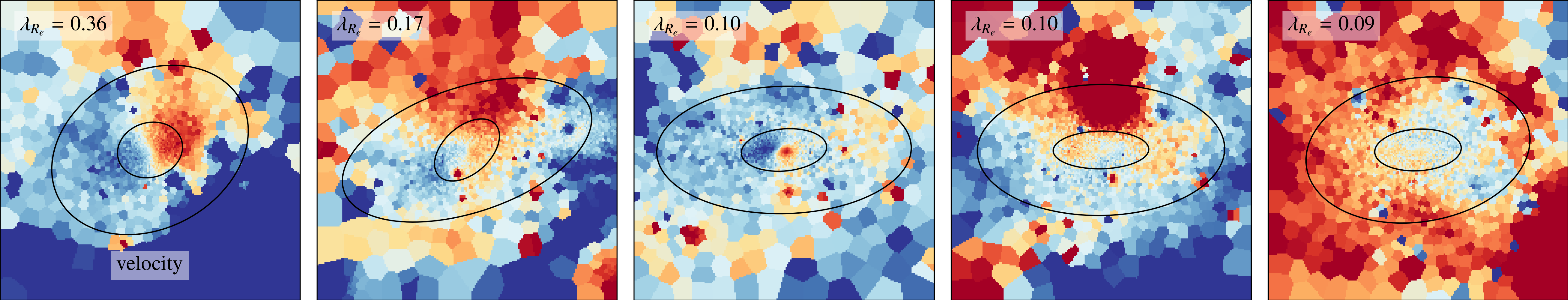}
    \includegraphics[width=\textwidth]{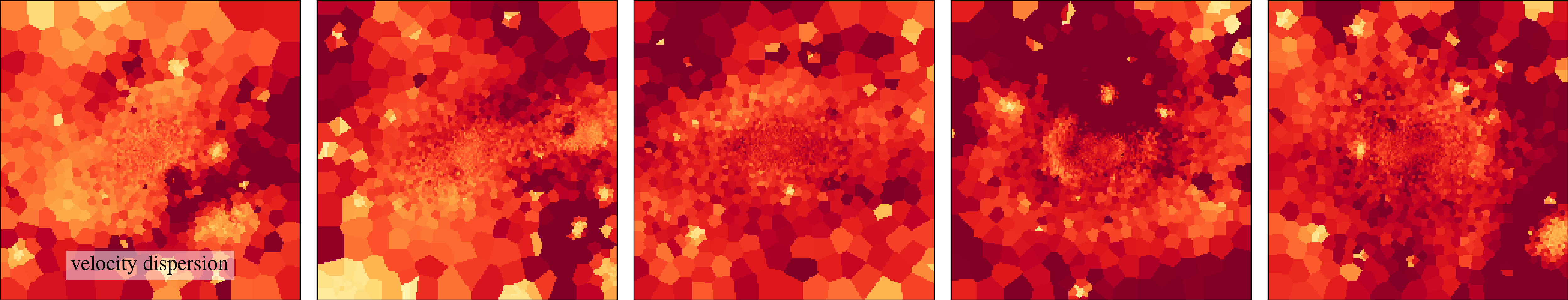}
    \includegraphics[width=\textwidth]{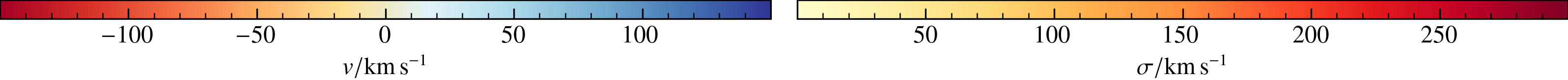}
    \caption{Evolution of the first example galaxy from \cref{fig:mocks}. The shells develop after a massive radial merger event, which also causes the value of $\lambda_{R_e}$ to sharply drop and converts the galaxy into a slow rotator (where $\lambda_{R_e}$ is measured from the edge-on projection). \textit{Top}: Mock images of the stellar component from a projection in which the shells are especially clearly visible. The red lines trace the orbit of the infalling galaxy until the respective point in time. \textit{Middle}: Velocity maps. The circles indicate the ellipses at one and three half-mass radii. \textit{Bottom}: Velocity dispersion maps.}
    \label{fig:evolution}
\end{figure*}

As an example, \cref{fig:evolution} shows the evolution of the most massive galaxy from our sample of galaxies with shells. The orbit of the infalling galaxy that triggered the formation of the shells is shown in red in the first row. The shells found around this galaxy were formed through a satellite with a stellar mass ratio of 1:2.5 falling in radially at $z = 0.20$ (almost \SI{2.5}{\giga\year} ago), then falling back in again on the other side at $z = 0.13$ (\SI{1.7}{\giga\year} ago). The back-and-forth movement led to the double shell structure found on two sides of the galaxy at $z = 0.07$ (\SI{0.9}{\giga\year} ago). This clearly supports the idea that shells form through mergers on radial orbits. In the second row of \cref{fig:evolution}, the velocity maps are shown. The system is transformed from a fast-rotating galaxy with $\lambda_{R_e} = 0.36$ to a slowly rotating galaxy with $\lambda_{R_e} = 0.09$ during the merger. Because these values are computed for the edge-on projection of the galaxy instead of the shown projection, the transformation into a slow rotator is physical. Finally, the last row shows the velocity dispersion, which overall increases through the merger and also contributes to decreasing the value of $\lambda_{R_e}$. Remarkably, the shells seen around the galaxy in the first row can also be clearly identified in the velocity dispersion maps, best seen in the map at $z=0.1$ as regions of comparably low velocity dispersion. This indicates that the kinematic studies of the outskirts of galaxies may be used to identify shells, for example, through tracer populations such as planetary nebulae \citep[e.g.,][]{roth+21}.

\section{Conclusion}
\label{sec:conclusion}

We classified galaxies from the Magneticum \MagneticumBox{4}{uhr} simulation with stellar masses above \SI{2e10}{\Msun} according to the existence of tidal features in their outskirts and used data of satellite planes from \citet{foerster+22} to connect their appearance to the internal kinematics of the host galaxies. We compared the results to those from \citet{bilek+20} and \citet{heesters+21} for the \matlas{} survey and observations of BCGs by \citet{tal+09} and \citet{kluge+20}. The internal kinematic properties were taken from \citet{schulze+18} and cover both overall rotational properties quantified by the $\lambda_R$ parameter \citep{emsellem+11:atlas3dIII} and internal kinematic patterns such as regular rotation, nonrotation, prolate rotation, and KDCs.

The fractions of Magneticum galaxies with shells, streams, and tails are comparable to those reported in observations and other simulations. Streams are found to be the most common of these features, with \SI{29}{\percent} of galaxies above a stellar mass of \SI{e11}{\Msun} exhibiting streams. We also see a clear trend with mass, with all types of features being more common around galaxies with higher stellar masses, in agreement with results from observations. Galaxies with more than one type of feature are also much more frequent at high stellar masses, indicating a much more turbulent accretion history for more massive galaxies. This agrees well with previous studies \citep[e.g.,][]{remus&forbes22}.

Additionally, we find indications that tidal features are more common around early-type galaxies than late-type galaxies for shells and streams. No shells were found for late-type galaxies, while tails are most frequently found for disk-like galaxies and are rather rare around spheroidal galaxies. This agrees well with the current scenario in which shells are formed during massive radial merger events that would also destroy the disk of a galaxy, while streams can occur around all types of galaxies as they are remnants of disrupted galaxies that fell in on circular orbits \citep[e.g.,][]{amorisco15,karademir+19}. Tidal tails, on the other hand, are best visible if a merger occurs to a disk galaxy because these galaxies are more radially extended at a given mass and their stars are therefore less strongly bound than those of spheroidal galaxies.

When inspecting the connection between global kinematic properties and the appearance of tidal features, we found a remarkable agreement with observations for the distribution of the kinematic parameter $\lambda_{R_e}$ for galaxies that have a given type of tidal feature. For both simulations and observations, slow rotators are most likely to have shells, while streams preferentially occur around galaxies with low to medium rotational support.
Tracing these galaxies with shells back in time, we find that \SI{30}{\percent} of the galaxies lose their global angular momentum through the merger event that also caused the shells. The remaining \SI{60}{\percent} of shell galaxies had already been slowly rotating prior to the merger event that created the shells. About half of these galaxies had lost their global angular momentum even prior to redshifts of $z=2$. Thus, we conclude that for all such cases where the galaxy was still rotationally supported, the massive merger event that caused the shells also led to a loss of the suppression of the rotational support. Moreover, we find that the orbits of these shell-forming satellite galaxies are radial, which agrees with previous simulation results from isolated merger experiments \citep[e.g.,][]{amorisco15,karademir+19} and from the Illustris simulations by \citet{pop+18}.

Interestingly, we also find that shells appear in the velocity dispersion maps around their host galaxies as regions with a low-velocity dispersion, while they do not appear in the velocity maps. While the faint outskirts of galaxies are unlikely to be probed by IFU surveys in the near future, kinematic tracers such as globular clusters and planetary nebulae can be used to map these regions and might therefore help detect shell features from their velocity dispersion distribution \citep[e.g.,][]{roth+21}. However, more detailed studies need to expand on these aspects in the future.

While we do not find a clear correlation between the overall rotational support of a galaxy and the appearance of streams beyond a slight tendency for streams to be the least common for galaxies with high rotational support, we find a connection between the appearance of streams and an internal kinematic feature: KDCs. For these galaxies, two different formation pathways are known: On the one hand, KDCs are made during gas-rich major mergers, with a lifetime of about \SI{4}{\giga\year} \citep{hoffman+10,schulze+17}. On the other hand, these KDCs can be remnants of an old disk that grew into a spheroidal through multiple minor and mini mergers on circular orbits that mostly populate the outskirts of these galaxies \citep{schulze+20}. About \SI{30}{\percent} of our KDC galaxies show stream features, the largest amount of any given tidal feature connected to a kinematic class. These clearly indicate that the formation pathway of these KDCs most likely corresponds to the multi-merger scenario.

The galaxies with the largest overall amount of features are prolate rotators for all types of tidal features. This agrees with observational results by \citet{ebrova+21} and with the fact that there are multiple possible pathways reported for prolate rotators \citep[i.e.,][]{ebrova&lokas17,hegde+22}, which also leads to a wide variety in possible tidal features. Overall, the large number of tidal features (sometimes even multiple different features) found around prolate rotators agrees with the idea that they are oftentimes formed through multiple merger events (effectively making them overmerged galaxies, as discussed by \citealp{remus&forbes22}), and thus, they are also rather frequent in high-mass galaxies.

Finally, we do not find any connection between the inner rotational support of the host galaxies and the existence of satellite planes. We find no strong indications of a connection between the appearance of tidal features and planes in addition to the possibility that tidal features may preferably occur around galaxies without satellite planes. Overall, satellite planes behave oppositely to the tidal features. They are most frequent around disk galaxies and regular rotators. This could either imply that the large-scale structure of accretion has no impact on the internal kinematics of a galaxy and the appearance of tidal features, or that these planes are the prestage of the merger events that leave the tidal features and change the kinematic properties of the host galaxy. Testing this in more detail will be an important step forward in the future.

\begin{acknowledgements}
We thank Pierre-Alain Duc, Elisabeth Sola, Klaus Dolag, and Lucas Kimmig for helpful discussions, and Nick Heesters for providing us with their data on the presence of satellite planes around \matlas{} galaxies, and the anonymous referee for useful comments that enhanced the quality of this manuscript.

We also thank Silvio Fortuné, Bendix Hagedorn, Elena Hernández Martínez, Tadziu Hoffmann, and Lucas Kimmig for participating in the classification process of the galaxies in the simulation.

LMV acknowledges support by the COMPLEX project from the European Research Council (ERC) under the European Union’s Horizon 2020 research and innovation program grant agreement ERC-2019-AdG 882679.

The calculations for the hydrodynamical simulations were carried out at the Leibniz Supercomputer Center (LRZ) under the project pr83li (Magneticum).

This research was supported by the Excellence Cluster ORIGINS, funded by the Deutsche Forschungsgemeinschaft under Germany's Excellence Strategy – EXC-2094-390783311.

The following software was used for this work: Julia \citep{bezanson+17:julia}, Matplotlib \citep{hunter07:matplotlib}, GadgetIO.jl \citep{boess&valenzuela:gadgetio.jl}.
\end{acknowledgements}

\bibliographystyle{style/aa}
\bibliography{bib}

\begin{appendix}

\section{Rescaling the cumulative distribution}
\label{app:rescaling}

The cumulative histogram of the simulated galaxies can be adapted to be directly comparable with that of the observations in the following way:
Qualitatively, the $\lambda_{R_e}$-values are divided into bins. Each bin contains a different total number of observed and simulated galaxies, as well as different numbers of observed and simulated galaxies with a given tidal feature. To make the latter numbers directly comparable to each other, the number of either the observed or the simulated galaxies with the given feature has to be multiplied with the ratio of the total numbers of galaxies in that $\lambda_{R_e}$ bin. In this work, we chose this to be the numbers of simulated galaxies with the given feature that are rescaled by the aforementioned ratio. The cumulative distribution is then obtained through summing over the individual bins up to a given value of $\lambda_{R_e}$.

Quantitatively, taking into account that there are different underlying distributions of values of $\lambda_{R_e}$ in the observations and the simulation and that the samples have different sizes, this means that for any interval $\Delta \lambda_{R_e}$, there is a different number of observed galaxies $N_\mathrm{obs}(\Delta \lambda_{R_e})$ and of simulated galaxies $N_\mathrm{sim}(\Delta \lambda_{R_e})$.
From these, the ratio of observed to simulated galaxies can be determined through $q(\Delta \lambda_{R_e}) = N_\mathrm{obs}(\Delta \lambda_{R_e}) / N_\mathrm{sim}(\Delta \lambda_{R_e})$. We used this fraction $q(\Delta \lambda_{R_e})$ to then make the number of simulated galaxies with a given feature that lie within such an interval directly comparable with the number of observed galaxies with that feature $N_\mathrm{obs,feat}(\Delta \lambda_{R_e})$.
For the $\lambda_{R_e}$ values of the simulated galaxies that have a given feature, $\lambda^1_{R_e} \leq \dotsb \leq \lambda^{N_\text{sim,feat}}_{R_e}$, where $N_\text{sim,feat}$ is the number of simulated galaxies with the feature, it is sufficient for cumulative histograms to choose the intervals $\Delta \lambda^1_{R_e} = [0; \lambda^1_{R_e}], \Delta \lambda^2_{R_e} = (\lambda^1_{R_e}, \lambda^2_{R_e}]$, etc.
This results in a set of $N_\text{sim,feat}$ intervals, for which the weightings of the $\lambda_{R_e}$ values of the simulated galaxies with the respective feature are given by the factors $q(\Delta \lambda_{R_e})$.
The sum over all these weightings then defines the rescaled number of simulated galaxies with the given feature, and the cumulative histogram can be determined through
\begin{equation}
    \frac{N_\mathrm{cum}({\leq} \lambda^i_{R_e})}{N_\mathrm{tot}} = \frac{\sum_{j=1}^i q(\Delta \lambda^j_{R_e})}{\sum_{j=1}^{N_\text{sim,feat}} q(\Delta \lambda^j_{R_e})}.
\end{equation}

\end{appendix}

\end{document}